\documentclass[lettersize,twoside,journal]{IEEEtran}
\usepackage{graphicx}
\usepackage{amssymb}
\usepackage{amsmath,cases}
\usepackage{cite}
\usepackage{multirow}
\usepackage{color}
\usepackage[table]{xcolor}
\usepackage{subfigure}
\usepackage{booktabs}
\usepackage{balance}
\usepackage{acro}
\usepackage{stfloats}

\usepackage{color}

\definecolor{darkgreen}{RGB}{0,128,0}

\newcommand{\M}[1]{\boldsymbol{#1}}

\newcommand{\B}[1]{\mathbf{#1}}
   
\DeclareAcronym{AO}{
short = AO,
long = alternate optimization
}

\DeclareAcronym{BS}{
	short = BS,
	long = base station
}

\DeclareAcronym{BD-RIS}{
short = BD-RIS,
long = beyond-diagonal reconfigurable intelligent surface
}

\DeclareAcronym{CSI}{
	short = CSI,
	long = channel state information
}

\DeclareAcronym{D2D}{
	short = D2D,
	long = device-to-device
}

\DeclareAcronym{FP}{
short = FP,
long = fractional programming
}

\DeclareAcronym{IAB}{
	short = IAB,
	long = integrated access and backhaul
}

\DeclareAcronym{IoT}{
	short = IoT,
	long = Internet of Things
}

\DeclareAcronym{IoD}{
	short = IoD,
	long = Internet of Drones
}

\DeclareAcronym{ISCC}{
	short = ISCC,
	long = Integrated sensing\, communication\, and computation
}

\DeclareAcronym{KPI}{
	short = KPI,
	long = Key performance indicator
}

\DeclareAcronym{LoS}{
	short = LoS,
	long = line-of-sight 
}

\DeclareAcronym{LSTM}{
	short = LSTM,
	long = long short term memory
}

\DeclareAcronym{mMTC}{
	short = mMTC,
	long = massive machine type communication
}

\DeclareAcronym{ML}{
	short = ML,
	long = machine learning
}

\DeclareAcronym{mmWave}{
	short = mmWave,
	long = millimeter wave
}

\DeclareAcronym{NTN}{
	short = NTN,
	long = non-terrestrial network
}

\DeclareAcronym{NW}{
	short = NW,
	long = Narrow band
}

\DeclareAcronym{nLoS}{
	short = NLoS,
	long = non line-of-sight
}

\DeclareAcronym{NLS}{
	short = NLS,
	long = non line-of-sight
}

\DeclareAcronym{PIN}{
	short = PIN,
	long = positive-intrinsic-negative
}

\DeclareAcronym{QNM}{
short = QNM,
long = quasi Newton method
}

\DeclareAcronym{RIS}{
short = RIS,
long = reconfigurable intelligent surface
}

\DeclareAcronym{RZF}{
short = RZF,
long = regularized zero forcing
}
\DeclareAcronym{SINR}{
	short = SINR,
	long = signal-to-interference-and-noise ratio
}

\DeclareAcronym{SNR}{
	short = SNR,
	long = signal-to-noise ratio
}

\DeclareAcronym{TN}{
	short = TN,
	long = terrestrial network
}

\DeclareAcronym{THz}{
	short = THz,
	long = Terahertz
}
\DeclareAcronym{URLLC}{
short = URLLC,
long = ultra-reliable and low-latency communication
}

\DeclareAcronym{UAV}{
	short = UAV,
	long = unmanned aerial vehicle
}

\DeclareAcronym{V2X}{
	short = V2X,
	long = vehicular-to-everything
}

\DeclareAcronym{WB}{
	short = WB,
	long = Wide band
}

\begin{document}

\title{
Quantum Intelligence Meets BD-RIS-Enabled AmBC: Challenges, Opportunities, and Practical Insights
}
\author{
Abd Ullah Khan, 
Uman Khalid, 
Trung~Q.~Duong,~\IEEEmembership{Fellow,~IEEE}, 
and
Hyundong~Shin,~\IEEEmembership{Fellow,~IEEE}
\thanks{
Abd~Ullah~Khan, 
Uman~Khalid,
and 
Hyundong~Shin (corresponding author)
are with the Department of Electronics and Information Convergence Engineering,
Kyung Hee University,
Yongin-si, Gyeonggi-do 17104,
Republic of Korea;
Trung~Q.~Duong 
is with the Department of Electrical and Computer Engineering, 
Memorial University of Newfoundland, 
St. John’s, NL A1B 3X5, Canada, and 
also with the Department of Electronics, Electrical Engineering, and Computer Science, 
Queen’s University Belfast, BT7 1NN Belfast, U.K.
}

}
\markboth{
Submitted for Publication in IEEE 
}
{ 
Khan \textit{\MakeLowercase{et al.}}:
Beyond-Diagonal \ac{RIS} for 6G networks: Principles, Challenges, and Quantum Horizons
}


\maketitle

\begin{abstract}
Ambient backscattering communication (AmBC) is deemed to be the game changer for the battery-less and zero-energy IoT communication.
Similarly, \ac{BD-RIS} is an advanced type of \ac{RIS} that has recently been proposed and is considered a revolutionary advancement in wave manipulation. 
At the same time, Quantum machine learning promises to revolutionize the existing networking paradigm by enabling many novel use-cases.
Coupled with Quantum machine learning, BD-RIS-enabled AmBC can potentially enable long range, and energy- and spectral-efficient IoT communication.
In this article, 
 we present recent advances in this domain and identify a series of challenges and opportunities.
To substantiate the proposed concept, we present a case study in which beamforming is designed for \ac{BD-RIS}-enabled AmBC and analyzed using four distinct algorithms. 
Additionally, the beamforming is enhanced through Quantum machine learning models
employing real-world communication scenarios. 
Consequently, we derive useful insights about the practical implications of Quantum-enhanced \ac{BD-RIS} for AmBC.
	
\end{abstract}



\acresetall

\section{Introduction}
\label{sec:1}



\IEEEPARstart{A}{mbient} backscattering communication (AmBC) is anticipated to revolutionizing IoT connectivity by enabling zero-energy communication and battery-less devices, promoting both energy and spectral efficiency.
It is estimated that over 125 billion IoT devices will be deployed globally by 2030. Enabling these devices with zero-energy communication capability will lead to both power, cost, and spectral efficiency. 
 AmBC-enabled devices can leverage existing RF signals in the environment by modulating and reflecting them for data transmission, thereby eliminating the need for a dedicated power source, transmitter, or active signal generation. 
This way, AmBC enables ultra-low-power, cost-effective, sustainable, and scalable communication networks, paving the way for smart cities, environment, industry, transportation, and healthcare.


Generally, AmBC consists of an RF source, a tag, and a reader. Tags 
are equipped with backscatter modules that modulate the incoming signal by reflecting or absorbing it. 
The modulation process involves varying the impedance of the device antenna to encode data. 
A receiver then picks up the backscattered signal and retrieve the transmitted data.

Due to resource-constrained, zero-energy, and passive communication nature, AmBC is limited to short-range communication and supports only small data-rate. 
To extend its coverage range, \ac{RIS} is being actively leveraged by the research community. 
An \ac{RIS} consists of an array of low-cost and energy-efficient passive reflecting elements that can be dynamically reconfigured to direct incoming wireless signals to desired locations by imposing specific phase shifts \cite{FM:24:IEEE_J_COML}. This capability allows \ac{RIS} to create virtual line-of-sight paths between transmitters and receivers, effectively overcoming propagation obstacles. 
 Being easily mountable on walls and buildings, \ac{RIS} allows network operators to manage wireless propagation environments and extend communication coverage, thereby facilitating massive connectivity. 

Although, the deployment of \ac{RIS} for AmBC-powered IoT has recently gained significant interest from the research community, traditional \ac{RIS} architectures face several performance limitations. 
For example, its configuration is limited in terms of passive beamforming manipulation, which can lead to sub-optimal performance in dynamic environments. Additionally, traditional \ac{RIS} reflects signals only in specific directions, blocking transmissions from passing through it and thereby limiting the coverage area of a network. These limitations arise from the simple architecture of \ac{RIS}, where each element acts independently and is directly connected to a load, resulting in a diagonal phase shift matrix.

The innovative concept of \ac{BD-RIS} emerges as a potential solution to address the limitations of traditional \ac{RIS} architectures \cite{LSC:23:IEEE_J_WCOM}. \ac{BD-RIS} aims to enhance flexibility, coverage, and performance by introducing more sophisticated configurations, resulting in a more general phase shift matrix. This provides greater flexibility in adjusting phase shifts, allowing for more dynamic adaptation to fluctuating wireless environments. \ac{BD-RIS} can also optimize coverage more effectively than traditional \ac{RIS} by controlling phase shifts in multiple directions, providing better coverage over larger areas.

Quantum-based network optimization has recently garnered significant attention from the research community due to its transformative potential across various aspects of communication \cite{wang2022quantum}.
 Specifically, radio access network, non-terrestrial networks, edge and fog computing are expected to be revolutionized by Quantum-based machine learning \cite{zaman2023quantum}. 
Given the inherent resource constraints in AmBC and the architectural complexity of BD-RIS, Quantum-based approaches can offer promising solutions.

This article presents an exhaustive exploration of \ac{BD-RIS}, aiming to highlight its various aspects regarding AmBC IoT.

This paper aims at exploring the potentialities of  BD-RIS-enabled AmBC systems. Additionally, it is aimed to explore the potentiality of Quantum-based approaches to tackle the complexities.

In this context, in Section \ref{sec:BD-RIS}, we describe \ac{BD-RIS} and its functional classification and architectural design. Subsequently, we present \ac{BD-RIS}-enabled AmBC in Section \ref{sec:BD-RIS-enabled-AmBC}. 
Next, the potentialities of Quantum-based approaches in this domain are discussed.
Lastly, a case study is presented to investigate its practical implications in Section \ref{sec:Challenges-n-Case-study}.

\begin{figure*}[t!]	
	\centering
	\includegraphics[width=\textwidth]{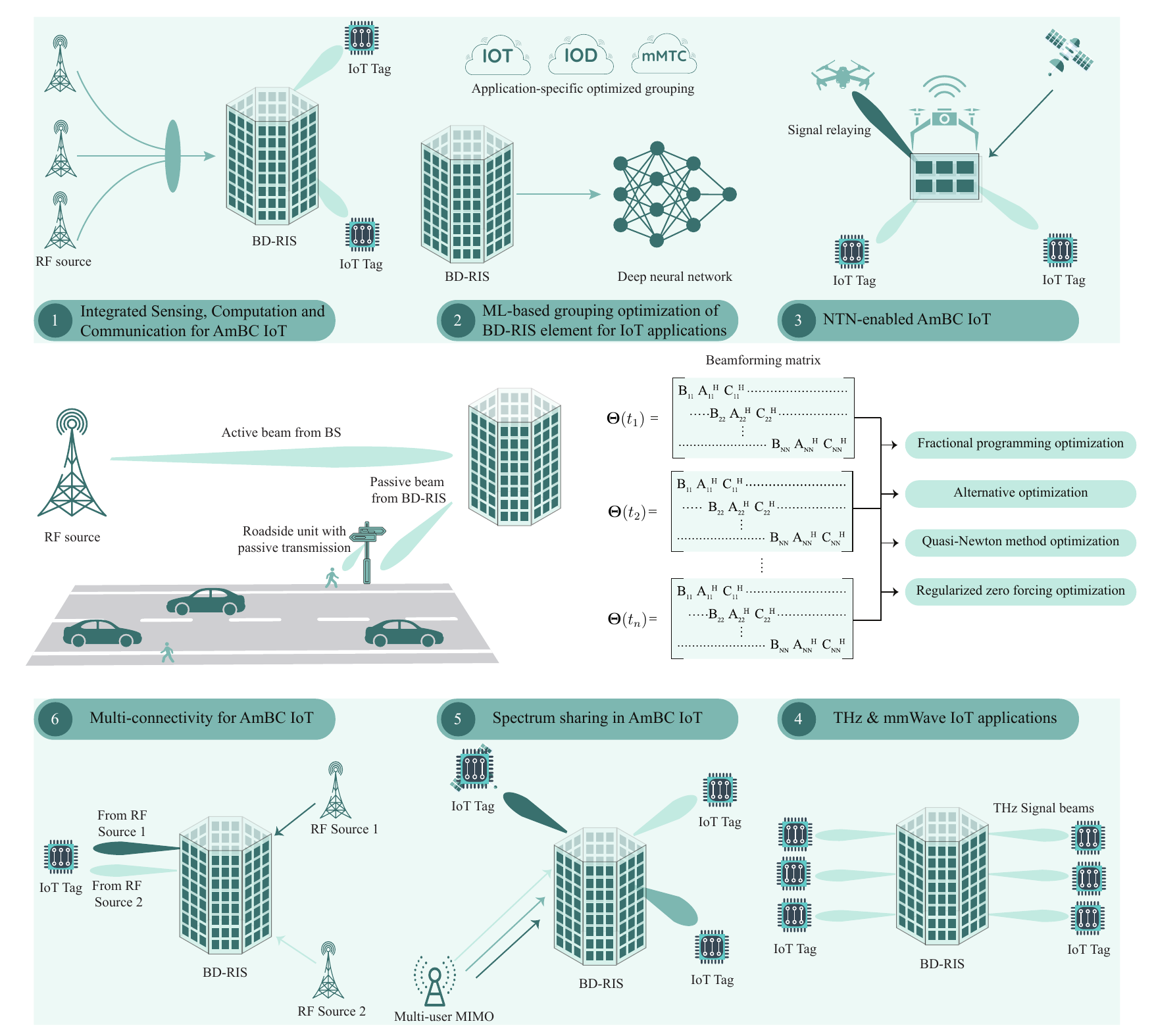}
	\caption{\ac{BD-RIS} for AmBC systems.}
	\label{fig:1}
\end{figure*}

\section{\ac{BD-RIS}} \label{sec:BD-RIS}

\subsection{Architectural Design}

\ac{BD-RIS} incorporates $N$ multiple passive elements acting as antennas linked to an $N$-port reconfigurable impedance network. This setup is characterized by a complex-valued $N \times N$ beamforming matrix $\M{\Theta}$, which not only describes the scattering properties of the network but also governs the interaction between the incident and reflected waves across $N$ ports. Moreover, the specific properties of the beamforming matrix depend on the employed circuit topology and allow for various architectural designs, each capable of non-diagonal configurations that augment \ac{BD-RIS}'s beamforming potential. 
Therefore, \ac{BD-RIS} fundamentally enhances the capabilities of RIS by incorporating an advanced reconfigurable impedance network, allowing wave manipulation with better precision and adaptability. In accordance with microwave theory, a lossless impedance network---represented by a unitary matrix---ensures that the power of the reflected waves matches that of the incident waves, optimizing energy efficiency and minimizing power loss. Such a design allows \ac{BD-RIS} to significantly empower wireless communication systems by offering dynamic control over wave propagation and interference.

\subsection{Functional Classification}

Functionally, \ac{BD-RIS} can be classified based on the characterization of beamforming matrices, operational modalities, and cell connectivity.

\subsubsection{Beamforming Matrix Characterization} 
\ac{BD-RIS} classification based on beamforming matrices is detailed as follows.

\begin{itemize}

\item 
\textbf{Block-Diagonal Matrix:} This type of beamforming matrix is formed when all $N$ antennas in \ac{BD-RIS} are uniformly divided into $K$ groups, with antennas within the same group are interconnected, while those from different groups remain disconnected. 
This configuration is referred to as group-connected, wherein each block in the beamforming matrix is unitary. 
However, when all antennas are interconnected ($K=1$), the group is termed full-connected, and the resulting beamforming matrix is also unitary under this arrangement. In contrast to the block-diagonal beamforming matrix typical of \ac{BD-RIS}, conventional \ac{RIS} features $K=N$ groups, wherein each group constitutes a diagonal beamforming matrix. This distinction underscores the enhanced reconfigurability for wave manipulation provided by \ac{BD-RIS} configurations.
	
\item 
\textbf{Permuted Block-Diagonal Matrix:} 
In configurations where the grouping of antennas is adaptively conducted considering the \ac{CSI}---also termed as dynamic group-connected \ac{BD-RIS}---the associated beamforming matrix is a permuted block-diagonal matrix. This offers higher flexibility in beam control compared to a fixed group-connected \ac{BD-RIS}, ultimately enabling adaptable manipulation of beamforming capabilities tailored to environmental fluctuations.
	
\item 
\textbf{Non-diagonal Matrix:} 
In this beamforming matrix configuration, antennas are paired through phase shifters, allowing the incident signal on one antenna to be reflected exclusively by its paired antenna. This results in an asymmetric non-diagonal beamforming matrix that achieves a higher power gain compared to conventional \ac{RIS}. This configuration improves the efficiency and effectiveness of signal transmission and reception by optimizing power distribution among the antenna pairs.

\end{itemize}

\subsubsection{Operational Modality} 

\ac{BD-RIS} classification regarding operational modes is itemized as follows. 

\begin{itemize}

\item
 \textbf{Reflective Mode:} 
In the traditional mode of \ac{BD-RIS}, the impinging signal is reflected back in the direction from which it came, effectively providing half-space coverage. In this architectural setup, all \ac{BD-RIS} antennas are oriented toward the same direction for uniform signal reflection. The associated beamforming matrix is subjected to a unitary constraint, characterizing the reflective mode of \ac{BD-RIS} operation. This constraint ensures that the matrix preserves the energy of the signal, which is crucial for maintaining the strength and quality of the impinging signal as it is reflected by the \ac{BD-RIS}.

\item 
\textbf{Tansmissive Mode:} 
In this operational mode, the impinging signal interacts with \ac{BD-RIS} and is transmitted to the other side, effectively extending network coverage for both the front and back sides of the \ac{BD-RIS}. Architecturally, this setup involves pairing two antennas back-to-back, enabling \ac{BD-RIS} to function not only as a reflective surface but also as a transmissive medium. The arrangement enhances \ac{BD-RIS} utility in complex networking environments by providing coverage on both sides.
	
\item \textbf{Hybrid Mode:} 
In this mode, the impinging signal is partially reflected by the \ac{BD-RIS} and partially transmitted to the other side, ensuring full-space coverage. Architecturally, this is achieved by placing antenna pairs back-to-back and connecting them through a 2-port fully-connected reconfigurable impedance network. To mathematically model this \ac{BD-RIS} arrangement, two matrices are required: the reflective mode matrix and the transmissive mode matrix.  This mode extends \ac{BD-RIS} functionality to multiple directions, such as 3-sided and 4-sided \ac{BD-RIS}, also termed multisector \ac{BD-RIS}. The gain and complexity of the system, as well as the number of matrices required for a comprehensive mathematical representation, significantly increase with the number of sectors. Therefore, \ac{BD-RIS}'s capacity to handle impinging signals from multiple directions is considerably enhanced by the multisector capability, making the hybrid mode particularly effective under intricate network conditions. 

\end{itemize}


\subsubsection{Cell Connectivity}

\ac{BD-RIS} classification concerning cell connectivity is listed as follows.
 
\begin{itemize}

\item 
\textbf{Single-connected:} 
This type represents the conventional \ac{BD-RIS}, where each antenna is connected to a single element. In such connectivity, the antennas operate independently and are isolated from one another, minimizing interference. The corresponding design is characterized by a diagonal beamforming matrix, where each diagonal element corresponds to an individual antenna's contribution without any cross-element interactions. The simplified formalism allows for a straightforward implementation and analysis but severely limits the \ac{BD-RIS} versatility in rather complex scenarios.

\item 
\textbf{Group-connected:} 
In this connectivity, antennas within the same group are interconnected, while those across groups are isolated. The beamforming matrix results in a block-diagonal matrix where each block corresponding to a group is unitary. This structure enables the strategic manipulation of incident waves---in both phase and magnitude---improving performance compared to single-connected \ac{BD-RIS}.
The grouping can be either fixed or dynamic, with the letter offering greater flexibility and performance enhancement at the expense of generating permuted block-diagonal matrices.

\item 
\textbf{Fully-connected:} 
In this configuration, all the antennas are interconnected, forming a fully-connected \ac{BD-RIS} that results in a unitary beamforming matrix.

\item 
\textbf{Tree-connected:} 
This category encompasses a family of \ac{BD-RIS} types that structurally form a tree graph when depicted using graph theory. It is comparatively less complex and performs equally well to fully-connected \ac{BD-RIS}. This tree graph connectivity enables efficient signal management while maintaining simplicity in network design.

\item 
\textbf{Forest-connected:} 
This is another family of \ac{BD-RIS} that gives rise to a forest graph when depicted using graph theory. It provides a balance in performance-complexity tradeoffs between the single-connected and tree-connected \ac{BD-RIS}, thus bridging the gap between them.

\end{itemize}

Among the listed \ac{BD-RIS} cell connectivity classes, the fully-connected and tree-connected architectures provide the highest flexibility and best performance. However, these architectures increase circuit complexity due to the large number of tunable impedance components. 
Alternatively, the group-connected and forest-connected architectures can simplify the circuit complexity but with reduced performance.

\subsection{Advantages} 
\ac{BD-RIS} offers various advantages compared to previous types of RIS. A few notable ones are listed as follows.

\subsubsection{High Freedom in Wave Manipulation}

By allowing the reconfiguration of both off-diagonal and on-diagonal entries in the beamforming matrix, \ac{BD-RIS} significantly enhances its ability to manipulate incoming signals. This flexibility results in improved performance gains by enabling precise control over the signal path and interaction within the matrix.

\subsubsection{Enhanced Network Coverage} 

By allowing adjustable configurations of both the impedance network and antenna array, \ac{BD-RIS} offers greater spatial coverage compared to \ac{RIS}. Specifically, the hybrid and multisector \ac{BD-RIS} modes enhance coverage by providing narrower beamwidths and higher antenna gains for each antenna, ultimately boosting performance in complex environments.

\subsubsection{Low-complexity Network Architecture} 
The hybrid and multisector variants of \ac{BD-RIS} offer enhanced deployment flexibility compared to the rigid deployment of traditional \ac{RIS}, facilitating comprehensive coverage across various spatial configurations. This flexibility simplifies network architecture, reducing complexity and improving space coverage.

\subsubsection{Cost-effective Network Deployment} 
The adaptable configuration and deployment of \ac{BD-RIS} leads to simple and cost-effective network architecture. Specifically, the fully-connected and group-connected reconfigurations offer optimized solutions for various network conditions, significantly reducing costs compared to conventional \ac{RIS}.

%
%
\section{Quantum Machine Learning}
6G aims to achieve over a ten-fold increase in spectral efficiency and more than a ten-fold improvement in energy efficiency. It supports a higher level of device connectivity, with a connection density of over $10^7$ devices per square km. It supports 3D wireless coverage in the form of TN-NTN convergence. It also enables massive connectivity by providing machine-type communication.
To meet the required data throughput, latency, reliability, and frequency utilization efficiency, 6G requires optimization of various parameters such as radio resource allocation, which involves laborious computation. Thus, quantum computing will enable efficient communication with the implementation of complex optimization processes.

Additionally, the interplay between the aforementioned enabling technologies enhances performance. For instance, RISs and non-orthogonal multiple access (NOMA) are jointly used to enhance spectral efficiency. However, such integration comes at a cost of significant increase in the computational complexity. 

Aside from this, wireless communication is expanding, giving rise to signaling and computational overhead. For instance, RISs and extra-large antenna arrays require a large number of pilot signals for channel estimation if conventional estimation techniques such as those relying on maximum-likelihood approaches are utilized.
Moreover, Considering the growing number of user terminals and nodes, it is essential to effectively allocate the limited network resources in real time.

AI and machine learning (ML) accelerated the transformation of wireless communication systems into natively intelligent systems capable of overcoming various challenges. However, the computational complexity of a classical-based learning model/algorithm generally grows with the dimension of the input data, e.g., channel state information, as well as with the complexity of the learning model, e.g., the number of layers composing the model, and with the number of iterations. This leads to an extended training duration and limits the applicability of high-dimensional learning models, especially for time-sensitive applications. 

The optimization in large scale wireless communication systems suffers from the challenge of large computational time and complexity. To overcome this challenge, quantum computing based algorithms are gaining attention. 
Quantum computing brings more powerful computation based on superposition and entanglement, and offers network optimization. 
For instance, in MIMO systems, quantum computing can optimize transmit power and enable smart beamforming optimization to maximize a user’s capacity and reduce latency with low power consumption. 
Similarly, IRS-assisted networks can be optimized using quantum computing. Specifically, the reflection coefficients can be efficiently optimized by quantum computing. 
Moreover, quantum computing can enable resource control mechanisms in IRS-assisted networks. 

Quantum computing also enables quantum machine learning, which can greatly promote ubiquitous wireless artificial intelligence (AI) in 6G as well. 
Realtime wireless AI is a new function or service in 6G that is enabled by Quantum machine learning. Both the AI training process and inference process are computation intensive, which can be expedited using quantum computing. This way, QML may introduce novel and more efficient wireless AI algorithms.

Quantum-based computations assume quantum bits, a.k.a. “qubits”, as the smallest unit of computation, each representing a superposition of computational bases enabled by a property called quantum superposition. Via the processing of multiple qubits, quantum-based computations can leverage other quantum properties such as quantum entanglement, which allows the state of a qubit to alter the state of another qubit, and quantum parallelism, which enables simultaneous information processing using a number of inter-connected qubits. 
Quantum-based learning models can attain faster training convergence. Quantum-based ML methods have also been shown to yield more accurate predictions compared to their classical- based counterparts. Similarly, QML uses less learning experiments to reach a specific optimization level.
The availability of general-purpose quantum processing units has supported recent studies on quantum-based ML.

Stated concisely, massive connectivity, rapid feedback, multiple dimensions, large intelligent surfaces, and real-time learning network states are the visions of 6G issues. An amalgamation of machine learning and quantum computing (QML) has been considered as kernel 6G enablers in wireless communication networks. Current prospects of quantum computing through discoveries on quantum mechanics such as inherent parallelism, more qubits integrated and quantum algorithms, clearly indicate a significant outperformance in terms of computational capability compared to conventional computing systems. Furthermore, supervised, unsupervised, and reinforcement learning in machine learning will enable big data analytic to assist self-organizing wireless networks based on the nature of data synthesis and the learning objective procedure. Consequently, joint quantum computing and machine learning is an appropriate solution for utilizing their joint benefits in the deployment of wireless systems. The parallelism concepts of qubit, entanglement, and superposition can handle huge data under large dimensional vectors in large spaces and generate statistical data patterns for machine learning methods.

\section{\ac{BD-RIS}-enabled AmBC for 6G IoT} \label{sec:BD-RIS-enabled-AmBC}

As outlined earlier, AmBC-enabled 6G IoT has limited communication range, and the research community is actively leveraging RIS to extend its communication distance. As RIS technology is constrained by an upper performance limit, \ac{BD-RIS} offers a promising solution for overcoming the limitations of RIS and enhancing achievable performance.
However, the technical research on \ac{BD-RIS} is still in its infancy, with only a few studies that explore its various aspects. Above this, no research has been conducted on its performance in the IoT environment. 
 In the following, we identify several challenges that must be overcome to unlock the true potential of \ac{BD-RIS} for 6G IoT.

\subsection{Challenges} \label{sec:Challenges-n-Case-study}

\subsubsection{Design Complexity}

Designing a practical prototype of \ac{BD-RIS}, especially in configurations such as group-connected with large groups and fully-connected, presents significant implementation challenges. Similarly, designing \ac{BD-RIS} in the hybrid mode with multisector capability involves inherent difficulties due to complex design interdependencies. 
Furthermore, the real-world deployment of \ac{BD-RIS} faces additional hurdles, such as losses in impedance components, mismatches in antenna setups, and mutual coupling of multiple antennas. These factors must be carefully accounted for in design and analysis to ensure proper functioning and efficiency in resource-constrained environment of AmBC-enabled IoT. Moreover, \ac{BD-RIS} using \ac{PIN}-based discrete values encounters problems when the beamforming matrix requires continuous values. Unlike traditional \ac{RIS}, where values can be quantized, the interdependent nature of the matrix quantities in \ac{BD-RIS} makes simple quantization strategies impractical. These complexities underscore the need for extensive research into various architectural designs and operational modes of \ac{BD-RIS}, tailored specifically to the resource-constrained environments characteristic of AmBC-enabled IoT systems.

\subsubsection{Realistic Channel Model}

The constraints associated with \ac{BD-RIS} hardware complexity significantly impact the development of channel models for these systems. The associated hardware challenges cause the channel models to become nonlinear functions of the beamforming matrix. This nonlinearity adds complexity to the beamforming design process for \ac{BD-RIS}, which can potentially reduce its performance in resource-constrained environment in AmBC-enabled IoT systems. The situation underscores the need for advanced research to simplify the hardware setup and manage the model complexities inherent in \ac{BD-RIS} systems.

\subsubsection{Channel State Information}

Like \ac{RIS}, \ac{BD-RIS} requires perfect \ac{CSI} for optimal beamforming and phase-shift adjustments. However, the prevalent methods used in \ac{RIS} to acquire \ac{CSI}---specifically the semi-passive and fully-passive methods---are either impracticable or too complex for implementation in \ac{BD-RIS}, particularly given power consumption constraints when deployed under AmBC-enabled IoT environment. Consequently, there is a need for the development of low-complexity techniques that can acquire \ac{CSI} in BD-RIS without significantly increasing power consumption.

\subsubsection{Inter-beam Interference in Multisector \ac{BD-RIS}}

When employing multisector \ac{BD-RIS} to generate multiple beams simultaneously from each sector, inter-beam interference becomes a critical issue. This interference can severely degrade network performance, with the severity intensifying as the number of sectors increases. Moreover, the increasing number of sectors not only enhances the network's capacity but also leads to a proportional increase in design complexity. This is particularly critical in resource-constrained environment of AmBC-enabled IoT. Consequently, meticulous design and deployment schemes are essential to leverage the potential benefits of \ac{BD-RIS} and effectively balance the complexity-capacity tradeoff. These schemes aim to optimize sector configurations to reduce inter-beam interference while increasing network throughput.

\subsubsection{Mobility Management}
 
Real-world IoT applications increasingly involve mobile devices such as \acp{UAV} and robots, rendering the static model irrelevant. Mobility introduces dynamic channel conditions, necessitating avant-grade approaches to tackle complex beamforming design, system modeling, and optimization to effectively address the fluctuating resource-constrained communication environment of AmBC-enabled IoT system.

\subsubsection{Computation Complexity and Sum-Rate Analysis} \label{sec:4:comp_complex}

The existing beamforming design schemes for \ac{BD-RIS} often rely on oversimplified scenarios for performance analysis, resulting in misrepresentation of its practical utility. In addition to the inherent complexities, these schemes face difficulties in convergence, leading to cost-ineffective realizations of \ac{BD-RIS}. Furthermore, their applicability becomes increasingly challenging when analyzing \ac{BD-RIS} in scenarios involving multiple (un)coordinated \ac{BD-RIS} with mobile devices.
This contributes to several problems that hinder the effective evaluation of \ac{BD-RIS} performance, particularly in terms of sum rate as well as computational cost analysis of \ac{BD-RIS}-empowered networks. This factor is particularly critical in resource-constrained environment such as AmBC-enabled IoT system.
Given these challenges, developing tactful approaches to designing \ac{BD-RIS} architectures that exhibit an optimal tradeoff between performance and computational efficiency with a focus on AmBC-enabled IoT system remains an open problem.

\subsubsection{Wideband Effects}

The frequency-dependent characteristics of \ac{BD-RIS} elements presents a challenge in design modeling and beamforming.
Due to impedance network circuits, the response of \ac{BD-RIS} elements varies with frequency, 
leading to the frequency response of each element being taken into account when modeling \ac{BD-RIS} for wideband applications. Contrarily, the interconnected configurations of \ac{BD-RIS} elements make element-wise modeling for wideband applications impractical.

\begin{figure}[t!]	
	\centering
	\includegraphics[width=0.5\textwidth]{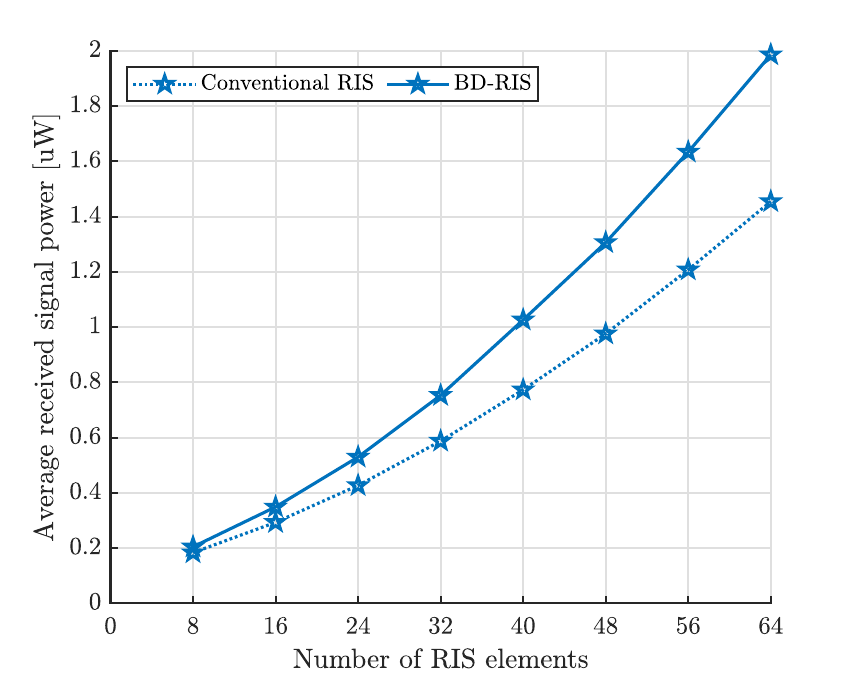}
	\caption{\ac{BD-RIS} and conventional RIS performance comparison for AmBC-enabled IoT.}
	\label{fig:RIS-vs-BD-RIS}
\end{figure}

\begin{figure*}[t!]
	\centering
	\subfigure[]{
		\includegraphics[width=0.49\textwidth]{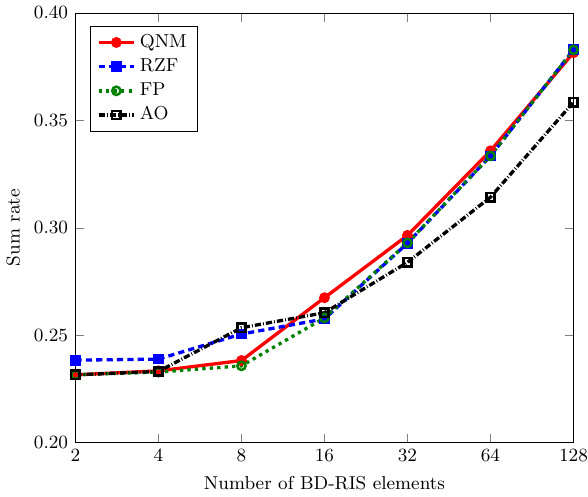}
		\label{fig:2a}
	}~
	\subfigure[]{
		\includegraphics[width=0.49\textwidth]{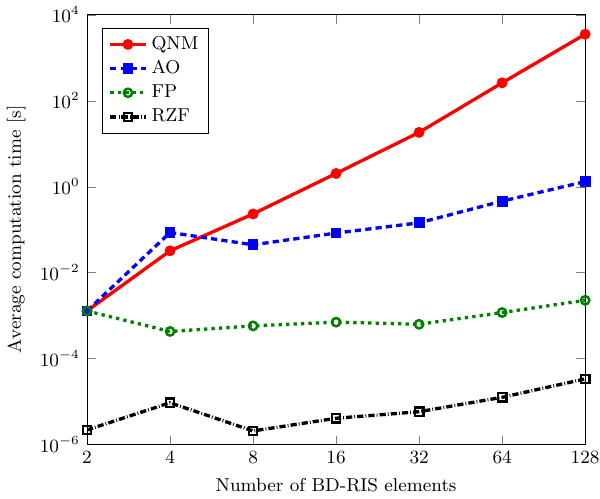}
		\label{fig:2b}
	}~
\caption{ 
Beamforming design performance. (a) The sum rate and (b) the average computation time are plotted as a function of the number of \ac{BD-RIS} elements for \ac{RZF}, \ac{FP}, \ac{AO}, and \ac{QNM}.
}
\label{fig:2}
\end{figure*}

\subsection{Case Study}

In the following, we present two case studies to demonstrate beamforming design for \ac{BD-RIS} in AmBC-enabled IoT and subsequently compare its performance with conventional approaches. We also elucidate how incorporating hybrid quantum-classical \ac{ML} in \ac{BD-RIS} could further advance its beamforming design capabilities.

\subsubsection{\ac{BD-RIS} for 6G IoT}

Here, we present a use case to analyze the performance comparison of \ac{BD-RIS} with conventional RIS for AmBC IoT. Additionally, we analyze computation cost and sum rate performance of four different algorithms for \ac{BD-RIS} beamforming design in AmBC IoT.

Fig. \ref{fig:RIS-vs-BD-RIS}, presents the average power received by a an IoT tag from a reflected signal by \ac{BD-RIS} and  conventional RIS. As the number of elements grows, the performance gap widens between the two RIS types.
Similarly, as mentioned, achieving an acceptable trade-off between computation cost and sum rate in beamforming design becomes critically challenging, especially under mobile and resource-constrained environment. In this context, we perform beamforming design using four different algorithms and analyze the associated computation cost and sum rate achieved for a device.
Specifically, we analyze a fully-connected \ac{BD-RIS} empowered multi-user multiple-input single-output system with device mobility. To this end, we consider an RF source, also called \ac{BS} for convenience, a total of $L$ devices spread over a location set $\mathcal{P}$, and \ac{BD-RIS} with a beamforming matrix $\M{\Theta}$. 
The channel matrices between devices and \ac{BS} are denoted by $\B{A}$, between \ac{BD-RIS} and devices by $\B{B}$, and between \ac{BS} and \ac{BD-RIS} by $\B{C}$, respectively. Specifically, we optimize the beamforming matrix for fully-connected \ac{BD-RIS} such that
$\max_{\M{\Theta}}\sum_{\ell=1}^L\sum_{p \in \mathcal{P}} \|\B{A}_{\ell,p}^\dagger+\B{B}_{\ell,p}^\dagger\M{\Theta}\B{C}\|^2$, where the superscript $\dagger$ denotes the conjugate transpose.
We design $\M{\Theta}$ and maximize the sum rate using \acf{FP} \cite{SY:18:IEEE_J_SP}, \acf{RZF} \cite{YTDPH:23:IEEE_J_WCOM}, manifold or \acf{AO} \cite{LSC:23:IEEE_J_WCOM}, and \acf{QNM} \cite{FM:24:IEEE_J_COML}.
Both small-scale and large-scale fading are assumed, with a reference path loss of $-30$ dB at a distance of $1$ meter (m). 
Path loss exponents are set at $3.5$, $2.2$, and $2.0$ for the channels between devices and \ac{BS}, devices and \ac{BD-RIS}, and \ac{BS} and \ac{BD-RIS}, respectively. The noise power is $-80$ dBm. Four antennas are assumed for \ac{BS}, and a transmit \ac{SNR} of $18$ dB is set. A random waypoint model is used for devices' mobility with the movement area restricted to $25 \times25~\text{m}^2$. Both line-of-sight and non-line-of-sight links are available to the end devices. The carrier frequency is $2.4$ GHz and the distance between \ac{BS} and \ac{BD-RIS} is $100$ m. 

Fig.~\ref{fig:2a} shows that \ac{QNM}, \ac{RZF}, \ac{FP}, and \ac{AO} perform increasingly well as the number of \ac{BD-RIS} elements increases.
Fig.~\ref{fig:2b} illustrates the computation cost for  \ac{QNM}, \ac{FP}, \ac{RZF}, and \ac{AO}. It can be seen that \ac{RZF} is the most computation-efficient, with its computation cost slightly increasing as the number of \ac{BD-RIS} elements grows.
In contrast, the computation cost of \ac{FP} remains nearly flat with an increase in the number of \ac{BD-RIS} elements. \ac{FP} performs better than \ac{AO} and \ac{QNM} but is less computation-efficient than \ac{RZF}, which is the most efficient option. Additionally, it can be observed that as the number of \ac{BD-RIS} elements increases, the computation cost of \ac{QNM}  drastically increases. Similarly, the computation cost of \ac{AO} also increases, though it remains lower than that of \ac{QNM}.

\begin{figure*}[t!]
	\centering
	\subfigure[]{
		\includegraphics[width=0.49\textwidth]{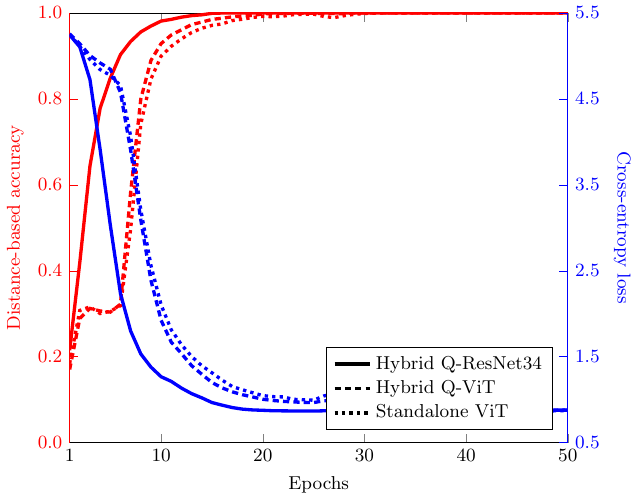}
		\label{fig:3a}
	}~
	\subfigure[]{
		\includegraphics[width=0.49\textwidth]{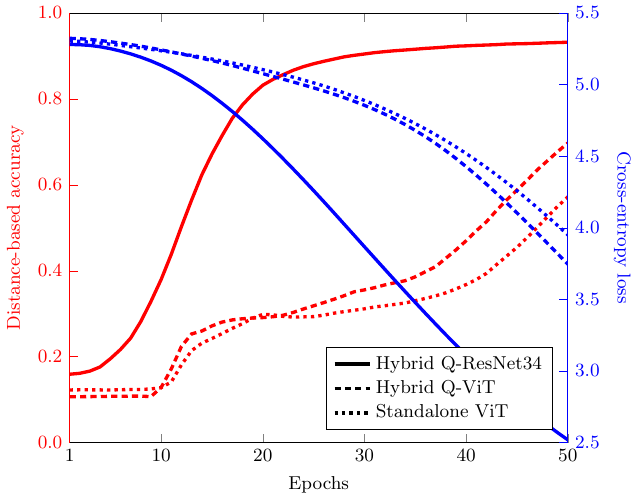}
		\label{fig:3b}
	}~
\caption{ 
Beam prediction performance: (a) The training and (b) validation results of distance-based accuracy and cross-entropy loss are plotted as a function of epochs for hybrid quantum-classical \ac{ML} models---i.e., quantum-ResNet (Q-ResNet34) and quantum-vision transformer (Q-ViT)---and a standalone classical \ac{ML} model---i.e., vision transformer (ViT).
}
\label{fig:3}
\end{figure*}

\subsubsection{Quantum \ac{ML} for \ac{BD-RIS}}

\begin{figure*}[t!]	
	\centering
	\includegraphics[width=\textwidth]{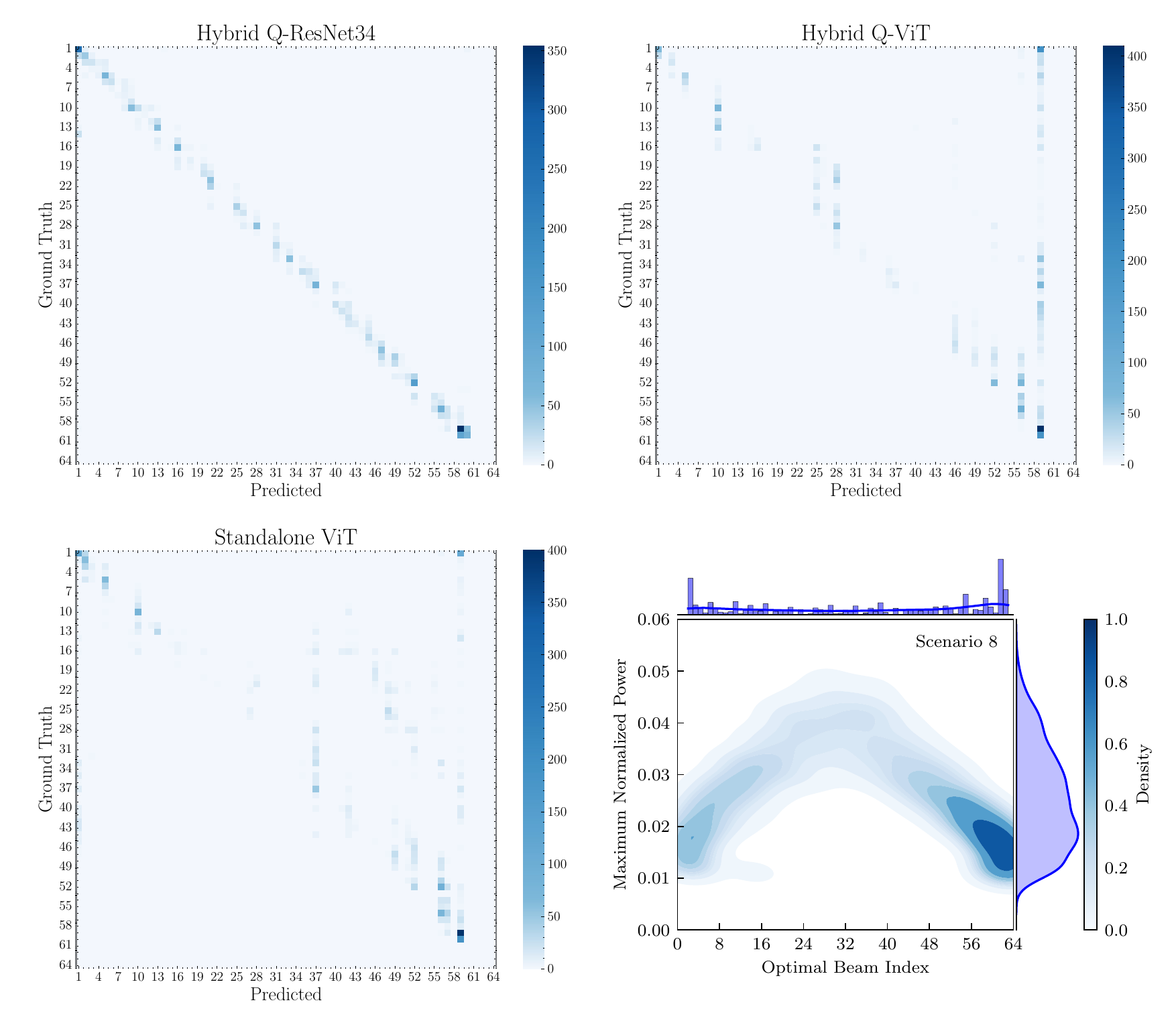}
	\caption{Confusion matrices for hybrid Q-ResNet34, hybrid Q-ViT, and standalone ViT. For Scenario~8 from the DeepSense 6G dataset, the frequency of each beam index and the variations in the distribution of the maximum normalized power are also demonstrated to identify any data bias that may lead to the deteriorated performance of these \ac{ML} models.}
	\label{fig:4}
\end{figure*}

Here, we present a use case to analyze the distance-based accuracy and cross-entropy loss for three different models, illustrating the advantage of employing hybrid quantum-classical \ac{ML} in a \ac{BD-RIS}-empowered AmBC-enabled IoT. Specifically, we evaluate the beam prediction performance using real-world communication Scenario $8$ from the DeepSense 6G dataset (RGB camera images and GPS location data). The parameterized quantum computing circuit is comprised of an amplitude embedding layer, an entangling layer, and a measurement layer. The output of the quantum model is concatenated with that of the classical \ac{ML} model for optimal beam prediction. As evident from Fig.~\ref{fig:3}, fusing quantum and classical \ac{ML} models---i.e., a hybrid quantum-vision transformer (Q-ViT)---can noticeably enhance beam prediction performance compared to its standalone classical counterpart. Moreover, integrating the quantum circuit with the ResNet34 model significantly outperforms the aforementioned models. Fig.~\ref{fig:4}  depicts the confusion matrices corresponding to the assessed \ac{ML} models and the distribution of optimal beam indices corresponding to the maximum normalized power and density estimates for Scenario 8 from the DeepSense 6G dataset.

\subsection{Applications and Opportunities} \label{sec:4:opport}

Following the numerical examples, it is evident that \ac{BD-RIS} can potentially increase data rate performance in AmBC IoT, and that a beamforming design algorithm with a striking balance between performance and complexity is crucial for \ac{BD-RIS} within AmBC IoT. Such algorithms not only enhance network efficiency but also pave the way to many novel applications and use cases for 6G IoT. Subsequently, we identify several applications and opportunities to augment the scope of \ac{BD-RIS} in 6G IoT.

\subsubsection{\acs{ISCC}} 

\Acf{ISCC} is an emerging technology that aims to combine these three elements, i.e., sensing, communication, and computation, into a single system, positioned as a key enabler for next-generation networks. It allows wireless signals to be used as sensors, such as for detecting obstacles. \ac{BD-RIS} enhances this integration by not only improving communication performance but also increasing the sensing accuracy, particularly for detecting devices in areas without line-of-sight. 
Therefore, \ac{BD-RIS} can potentially empower such integrated systems specifically within an AmBC-enabled IoT architecture that is both cost-effective and simplified.

\subsubsection{\ac{ML}-based Optimization}

The prevailing research mostly considers static grouping strategies for \ac{BD-RIS} elements, leading to sub-optimal performance in real-world scenarios. To address this shortcoming, real-time data-driven grouping strategies should be employed, incorporating dynamic factors such as \ac{CSI} and application-specific requirements. Alternatively, grouping strategies can be tailored for applications like the AmBC-enabled IoT, \ac{mMTC}, and the \ac{IoD} by leveraging quantum, classical, or hybrid \ac{ML}-based optimization approaches to enhance grouping efficiency \cite{ZFUJSW:23:IEEE_M_WC}. In particular, \ac{LSTM} networks can be highly effective in optimizing the grouping of elements in \ac{BD-RIS} configurations, thereby improving the formation of the beamforming matrix and overall system performance.

\subsubsection{Devices' Mobility}

Incorporating a dynamic environment into \ac{BD-RIS}-based AmBC-enabled IoT networks brings both challenges and opportunities. An intriguing application in this direction is integrating \acp{UAV} with \ac{BD-RIS}, enabling them to serve as both mobile source and relays. \acp{UAV} are increasingly recognized as an effective technology for use as mobile source. By equipping \acp{UAV} with \ac{BD-RIS} capabilities, numerous novel use cases could be formulated, especially in the context of \acp{NTN}-enabled IoT. Such integration fosters communication flexibility and coverage in response to dynamic environmental fluctuations in remote areas based IoT applications, thus broadening the operational scope of AmBC-enabled IoT systems limited by mobility constraints.

\subsubsection{\ac{THz} and \ac{mmWave}}

Owing to the multidimensional coverage capabilities, \ac{BD-RIS} is potentially effective at both \ac{mmWave} and \ac{THz} frequencies, where the high-directivity gain is required to mitigate path loss and enlarge the coverage area. 
The research gap concerning this open problem presents a significant opportunity to investigate beamforming design optimization for \ac{mmWave} and \ac{THz} channels in AmBC-enabled IoT systems, develop efficient \ac{CSI} acquisition and feedback mechanisms with corresponding overhead analysis, and explore \ac{BD-RIS} hardware limitations while operating at these frequencies in resource-constrained environment.

\subsubsection{\ac{BD-RIS} for \acp{NTN}}

The integration of \acp{NTN}-based IoT has recently attracted much interest from the research community, leading to many proposed use cases leveraging \ac{RIS}-enabled \acp{NTN}. However, the exploration of \ac{BD-RIS}, particularly with multisector and hybrid modes, remains largely unexplored. In this context, a promising research direction is to explore channel modeling for \ac{BD-RIS}-enabled \ac{NTN}-based IoT integration.

\subsubsection{Spectrum Sharing}

Hybrid-mode \ac{BD-RIS} can potentially revolutionize the spectrum-sharing paradigm by enabling numerous use cases and applications including AmBC-enabled IoT.
In particular, the group-connected and multisector \ac{BD-RIS} operating in hybrid mode can effectively promote spectrum sharing between incumbent and secondary devices in symbiotic radio enabled IoT, thereby mitigating interference effects. In other words, \ac{BD-RIS} has the potential to further advance spectrum-sharing capabilities in resource-constrained IoT environment by building upon the foundations laid by \ac{RIS} in optimizing spectrum utilization.


\subsubsection{Multi-connectivity and Multi-beam Support}

\ac{BD-RIS} can potentially revolutionize multi-connectivity and multi-beam support by leveraging its enhanced control over the electromagnetic properties of signals, thereby directing multiple beams to simultaneously serve multiple devices in 6G IoT. This is particularly important in cases where various data packets need to be directed to devices across various locations. 
In this way, \ac{BD-RIS} can play a crucial role in enhancing reliable and and low-latency requirements of 6G IoT.


\balance

\section{Conclusion}
\label{sec:5}

AmBC-enabled IoT has emerged as the energy- and spectral-efficient solution for 6G IoT. Similarly, the emergence of \ac{BD-RIS} marks a significant advancement in \ac{RIS} technology. 
 This article reviews the \ac{BD-RIS} functionalities and its applications in AmBC-enabled IoT. To this end, this article identifies the core challenges, highlights the key advantages, and exemplifies the complexity-performance tradeoff and quantum/classical \ac{ML}-based performance enhancement through relevant case studies. Additionally, this article also facilitates \ac{BD-RIS} applicability in 6G IoT by presenting novel applications and pointing out further research opportunities, thereby catalyzing its integration into 6G IoT.



\bibliographystyle{IEEEtran}
\bibliography{IEEEabrv, ./LatexInclusion/IEEE-Network-BDRIS}

\end{document}